\documentclass[12pt,preprint]{aastex}

\shorttitle{Detection Rates for Transits}
\shortauthors{T.M. Brown}

\begin{document}

\title{Expected Detection and False Alarm Rates for Transiting Jovian Planets}

\author{Timothy M. Brown}
\affil{High Altitude Observatory/National Center for Atmospheric Research,
Boulder, CO 80303}

\email{timbrown@hao.ucar.edu}

\begin{abstract}
Ground-based searches for transiting Jupiter-sized planets have so far produced
few detections of planets, but many of stellar systems with eclipse depths,
durations, and orbital periods that resemble those expected from planets.
The detection rates prove to be consistent with our present knowledge
of binary and multiple-star systems, and of Jovian-mass extrasolar planets.
Space-based searches for transiting Earth-sized planets will be
largely unaffected by the false alarm sources that afflict
ground-based searches, except for distant eclipsing binaries
whose light is strongly diluted by that of a foreground star.
A byproduct of the rate estimation is evidence that the period
distribution of extrasolar planets is depressed for periods between
5 and 200 days.
\end{abstract}

\keywords{stars: planetary systems, stars: binaries: eclipsing, techniques: 
photometric}

\section{Introduction}

Efforts to detect transits by Jovian-mass planets in tight orbits about Sun-like
stars began soon after the detection of 51 Peg b \citep{may95},
and have intensified
to the point
that there are now some 20 such transit searches underway \citep{hor02}.
Results from these transit searches have been disappointing, however.
For example, the Vulcan project \citep{jen02} searched 
6000 stars, locating 7 that displayed the shallow
and short-duration eclipses that would characterize a planet.
But spectroscopic and photometric follow-up observations showed that
all of these eclipses result from stellar, not planetary, systems.
Similar results from the STARE instrument \citep{bro99} will be reported
elsewhere.
The OGLE-III survey (Udalski et al. 2002, 2002a) 
imaged some $5 \times 10^6$ stars, 
of which
52000 were selected as likely main-sequence dwarfs. 
Of these, 59
candidates showed eclipses that appeared consistent with planetary transits.
Of the 59 transit candidates, 6 are too faint for follow-up observations,
and 3 may be planets \citep{kon03,dre03},
although the evidence for these is not yet conclusive.

Why then do ground-based searches yield so many false alarms and so
few planets?
More generally, can one understand the observed detection rates for
planets and for stellar systems that masquerade as planets, given
existing knowledge about stars, multiple-star systems, and extrasolar
planets?
Finally, what does this understanding imply for the ambitious
spaceborne missions COROT \citep{bag00}, Kepler \citep{koc98} and Eddington
\citep{gim01}, which aim to use transits to detect
Earth-sized planets in Earth-like orbits?

\section{Estimating Detection Rates}

Estimating transit detection rates is complicated because many different kinds
of stellar systems can produce transits resembling those from planets,
and each of these requires its own methods, assumptions, and approximations.
I categorize these cases in a tree structure, defined by the
answers to the following questions:
(1) Is the primary of the eclipsing system 
a main-sequence star (M) or a giant (G)?
(2) Is the secondary a main-sequence star (S) or a Jovian planet (P)?
(3) Is the light from the binary undiluted by a third star (U), or
diluted (D)?
(4) If diluted, is the third star a foreground object (F) or is the system
a bound triple (T)?
This scheme yields 12 possibilities.
These classes are by no means exhaustive, but they encompass
the kinds of systems that are most commonly found in ground-based searches.

Of the 12 classes, I shall discuss only 5.
The targets of ground-based searches are denoted MPU, ie, the primary
is a main-sequence star and the secondary is a giant planet.
The principal sources of false alarms involve grazing transits in systems
of 2 main-sequence stars (MSU), diluted instances of these (MSDF, MSDT),
and transits of giants by main-sequence stars (GSU).
The remaining categories are less important because they are less common,
because their eclipses do not usually simulate planetary
transits, or both.

To estimate detection rates,
one wishes to know the probability of finding objects of each class
with photometric properties lying within 
ranges that are appropriate to planetary transits.
These ``observable'' photometric properties are the transit repetition period
$\Pi$ (days), the transit duration $d$ (days), and the relative 
depth of the transit $\delta$ (in the range $\{0,1\}$).
In addition, one can usually say something about the nature of the
primary star;  for main-sequence stars one may think of this as the
primary's mass $m_0$ (solar units), while for giants it is more convenient
to use the J-K color $c_0$, measured in magnitudes.
The joint probability density (for instance, $P(m_0, \Pi, \delta, d)$)
then describes
the probability that a randomly-chosen target star in a photometric
survey is a main-sequence eclipsing binary with properties near 
the given ones.
To obtain an estimate of the number of detections to be expected
in the target sample from this class of object,
one integrates this density over the interesting ranges of
observable parameters and multiplies by the number of target stars
in the sample.
Marginal probability densities 
(integrated over all parameters but one) are also
useful, and some of these will be displayed below.

Unfortunately, the joint probability density in terms of the observable
parameters is not ordinarily available.
Instead, one has empirical estimates (e.g., \citet{duq91}, henceforth called
DM) 
of probability densities for
``physical'' parameters taken singly, such as $P(\Pi)$, $P(m_0)$, $P(\cos i)$, and
$P(q)$, where $i$ is the orbital inclination to the line of sight, and
$q$ is the
binary mass ratio. 
To make these useful, one must first assume independence among the various
physical parameters, so that the joint density may be written as a product
of the individual densities.
One must then transform the density $P(m_0, \Pi, q, \cos i)$
into a density on the observable parameters $P(m_0, \Pi, \delta, d)$.
Details of this process for the various cases will be published elsewhere;  
for current purposes, all that matters is that the transformations
contain no essential 
singularities.
More important are the empirical distributions, the relations among stellar
properties, and the assumptions invoked to derive the 
initial joint distributions and the transformations.
These are described below.

I characterized main-sequence binaries (type MSU) by the 4 physical 
parameters discussed above,
viz., $\{ m_0, \Pi, q, \cos i \}$.
I estimated $P(m_0)$ using the local main-sequence luminosity function
$N(M_V)$ giving the space density of stars as a function of their V absolute
magnitude, the mass-luminosity relation $M_V(m_0)$, and color-color
relations (e.g. $M_R(M_V)$, where $M_R$ is the absolute magnitude for
R-band photometry).
I obtained all of these relations from the tables in \citet{cox00},
except that I adjusted the luminosity function to provide better
agreement with the observed histogram of J-K color, for a field lying on the
galactic plane in Cygnus, overlapping the proposed field for the Kepler
mission.
Because of interstellar extinction,
calculating $P(m_0)$ also requires knowledge of the
photometric bandpass and limiting magnitude appropriate to the survey
being modeled, and of $D_e$, defined as the typical distance
needed to accumulate 1 magnitude of interstellar extinction in the V band.
I used $D_e = 1000$ pc, and assumed that all observations are in the
galactic plane.
I took $P(\Pi)$ and $P(q)$ from DM, and assumed
the fraction of spatially unresolved main-sequence stars having 2 or more
components to be 0.49, in accord with their estimate.
I assumed orbital axes to be randomly oriented in space, giving
$P(\cos i) = 1$.
Finally, I adopted several simplifications to facilitate the calculations.
The most important of these were: (1) I ignored the luminosity of the
secondary component, relative to that of the primary.
(2) I simplified the mass-radius relation on the main sequence, taking
radius (in solar units) to be numerically equal to mass (in solar units).
(3) I took the stellar limb-darkening coefficient to be $u=0.5$,
irrespective of wavelength or stellar mass.
(4) For purposes of computing the system's semimajor axis, I ignored
the mass of the secondary component.
(5) I assumed all binary orbits are circular.
These approximations limit the accuracy of the final result;
combined with the uncertainties in the input probability
densities, they suggest that the final estimates of detection rates
are likely to be inaccurate by factors of a few.

For binaries with one giant component (type GSU), I took the same distributions
$P(\Pi)$, $P(\cos i)$ as for main-sequence binaries, except
that I set $P(\Pi)$ to zero for periods such that the secondary
would lie within the surface of the primary.
I parameterized
the giant primaries in terms of their V-K colors,
by inventing an analytic 3-component distribution 
$N(\rm {V-K})$. 
This, combined
with the relation (for giants) between spectral type and V-K from
\citet{cox00}, gave space densities for spectral types G, K, and M;
I chose parameters in the analytic distribution to make these
consistent with those tabulated in
\citet{mih81}.
Given the V-K color, I computed the giant radius and luminosity from
relations given by \citet{vanB99}.
Color-color relations from \citet{cox00} then allowed estimation of
$P(c_0)$.
I assumed the distribution of secondary masses to be the same as 
$P(m_0)$, the distribution of main-sequence primary masses.

I assumed that main-sequence binaries that are diluted by foreground stars
(type MSDF) consist of a binary with properties drawn from the MSU distribution,
superposed on a foreground star whose R magnitude lies within
the range of R magnitudes covered by the survey being modeled. 
I used the tabulation of
number of stars $N(V)$ per V magnitude per square degree of sky taken from
\citet{all73}, for a galactic latitude of 0$^\circ$
(scaled to match the counts for $9 \leq \rm R \leq 12$ observed in Cygnus
by the STARE telescopes)
to estimate the number of background stars in each interval
of magnitude lying within a specified confusion radius (in arcsec)
of the foreground
star.
I then computed the distribution of apparent transit depth from 
the distribution of relative brightnesses
of the foreground and background stars, combined with the MSU distribution
of transit depth for the background system taken alone.

I assumed
that triples (type MSDT) are composed of a main-sequence binary drawn from the
MSU distribution, combined with a third star drawn from the observed
luminosity distribution.
This 
ignores triples in which 2 or more
members are giants.
Calculation of the relative depth distribution then proceeded in the
same way as for MSDF systems.

The probability distribution for planets orbiting main-sequence stars
(type MPU)
may be calculated by methods exactly analogous to those for MSU
systems, except that the DM distributions for
period and for secondary mass (implying radius, in the stellar case)
are inapplicable.
I estimated the distribution of Jovian planetary periods $P(\Pi_p)$
using the tabulation by
\citet{sch03}.
Figure 1 shows the normalized cumulative distribution 
of $\Pi_p$ against $\log(\Pi_p)$
for the 103 planets listed as of 4 April 2003;
the derivative of this distribution is the probability density
$P(\log \Pi_p)$, for stars having at least one Jovian planet.
Also shown is a piecewise linear fit to the distribution, corresponding
to $P(\log \Pi_p) \ = \ \{.509, .165, .533 \}$ for the 3 period intervals 
bounded by $\log \Pi_p \ = \ \{.43, .65, 2.3, 3.5\}$.
Multiplying this by the probability that a given star has a Jovian
planet, which I took to be .05 \citep{mar00}, gives the probability density that 
I assumed for calculating the MPU
distributions.
The relative dearth of planets with periods between 5 and 200 days
is a striking feature of this density.
Udry, Mayor, \& Santos (2003) have remarked on it also, arguing that
it carries information about the
migration (or, less likely, formation) process that places
massive planets in the orbits they now occupy.

In the interest of simplicity, I assumed that the distribution of planet
radii is uniform between 0.08 and 0.15 solar radii (0.78 to 1.46
Jupiter radii).
This spans the range from objects slightly smaller than Saturn to ones
slightly larger than HD 209458b.
I ignored the likely relation between a planet's equilibrium
temperature and its radius.

Finally, one must consider
the probability that transits with a particular period
will be observed, given a realistic set of observing epochs.
The observing window is a major concern for ground-based observations,
but will presumably be less important for those from spacecraft.
Except in cases with unusually low noise, detections of
single transits are almost useless (because many processes make
false signals that look like isolated transits), and 2-transit detections
are suspect.
The most reliable cases are those with 3 or more transits.
For a typical observing run using the STARE telescope 
(211 hours of observation on 38 individual nights, spread over an interval
of 91 days),
even for periods as short as 3 days, the probability
of seeing 3 transits with the observations just described is only about
0.5.
For longer periods, the visibility drops sharply, accentuating the
already strong bias towards short-period planets.

\section{Detection Rates for $\rm R \leq 12$ and $\delta \geq .01$}

Based on the assumptions described in the previous sections, I
computed detection rates for Jovian planets and for the 4 false alarm
categories, for a survey with properties similar to Vulcan or STARE.
To do this, I integrated the probability densities over all stellar
types, over periods in the range $1 \rm d \leq \Pi \leq 30$d,
over relative depths $.01 \leq \delta \leq .05$, and over transit
durations $.06 \rm d \leq d \leq .25$d.
(The lower limit on $d$ comes from detectability considerations, but few
additional transits would be counted if it were lowered to zero.)
I assumed a confusion radius of 20 arcsec (about 1.8 detector pixel
widths, for these systems), and a limiting R magnitude of 12.
The leftmost 3 columns of Table 1 show the estimated 
detection rates per $10^4$ stars
(this being approximately how many stars one can observe simultaneously
with current facilities),
for detection of at least 0, 2, and 3 transits, assuming the observing
window function described in the previous section.

The most important result from Table 1 is the low rate of planet
detection -- only about 0.4 detection per $10^4$ stars, for 3-transit
detections.
Naive estimates give detection rates about 5 times larger than this.
Nevertheless, the value from Table 1 is probably an overestimate, 
since a proportion
of main-sequence stars are members of close binaries in which small
planetary orbits would be unstable,
and since image crowding and other effects make it difficult to
achieve 1\% photometric precision for all target stars.

There are two principal reasons for the small expected rate of planet
detection.
The first is the low efficiency of single-site observations, as already
discussed.
The second is best illustrated by using the luminosity function and
color-color relations to construct a histogram showing the expected
number of observed stars per square degree as a function of J-K color.
Figure 2 shows this histogram, as calculated from the various model
relations and also as observed for the STARE Cygnus field, using the 
Two Micron All Sky Survey (2MASS)
catalog as the source of J-K colors.
In Fig. 2, stars with J-K $\leq 0.35$ are almost exclusively dwarfs,
while those with J-K $\geq 0.5$ are predominantly giants.
The hatched area, containing only 14\% of all stars, shows those
stars with radii less than 1.3 $R_\sun$,
i.e., those for which a central transit by a fairly large planet would
have $\delta \geq .01$.
Thus, even among main-sequence stars, only about a third of the stars
in the sample will have radii small enough for transits to be observed
in a routine way.
Moreover, any failings in photometric precision cause a disproportionate
decrease in the planet catch.
Figure 3 shows the expected distribution of $\delta$ for transits by
planets, by main-sequence binaries, and by diluted binary and triple systems.
Only about a third of all planetary transits should have $\delta$ as
large as .01, and the fraction drops rapidly as the $\delta$ threshold
is raised.

The most common source of false alarms is main-sequence
binaries (type MSU).
These should occur at a rate of several per $10^4$ target stars, and
should typically show relatively large depths and short periods.
Diluted main-sequence transits account for the rest of the false alarms,
about equally distributed between types MSDF and MSDT.
Although they are intrinsically rare, the triple systems appear fairly
commonly in transit searches because their $\delta$ distribution
peaks near .01, the canonical value for transits by Jovian planets.
Integrating over the full ranges in all of the parameters \{ $m_0$, 
$\Pi$, $\delta$,
$d$ \} shows that about 1.2\% of all stars observed should reveal
themselves as eclipsing binaries (ignoring window function effects).
Although detailed comparisons with these predictions have not yet been
made, these false alarm and eclipsing-binary rates seem to agree with
experience from the STARE and Vulcan surveys, at least within a factor
of 2.

\section{Detection Rates for $\rm R \leq 14$ and $\delta \geq 10^{-4}$}

Space missions targeting Earth-sized planets in Earth-like orbits
will search a different range of parameter space, seeking shallower
transits by objects in longer-period orbits.
For the purposes of these missions, all of the signal categories so
far considered represent possible false alarms.
The rightmost column of Table 1 shows their expected rates of 
occurrence per $10^4$ target stars,
for a survey that is magnitude-limited at $\rm R = 14$, and for
transit parameters in the ranges $30\rm d \leq \Pi \leq 300$d,
$5 \times 10^{-5} \leq \delta \leq .001$, and $.2 \rm d \leq d \leq 1.0$d.
The MSDF value assumes a confusion radius of 6 arcsec (about 3 pixels
widths for the Kepler instrument), and I assumed that 
the observing window function plays no role.
Note that the MPU entry in this column refers only to false alarms 
from grazing transits by Jupiter-sized planets, not to true detections
of Earth-sized planets;
if all Sun-like stars have solar systems similar to our own, the number of
true detections should be $\sim 10$ per $10^4$ target stars.

By far the largest source of false alarms in this case is faint background
main-sequence binaries.
The number of such binaries per square degree of sky increases rapidly
as the binaries grow fainter, so that the marginal distribution $P(\delta)$
rises sharply for small $\delta$.
Although false alarms of type MSDF will be common in spaceborne surveys,
they will be easy to identify and ignore.
The easiest discriminant
will be to compare the centroid of the target star image with
the centroid of the transiting signal.
Displacements of a fraction of a pixel can be detected in this way,
filtering out more than 99\% of the background binaries.

\section{Discussion}

Results from searches for transiting Jovian planets
can be understood in terms of
our existing knowledge of binary star systems and of extrasolar planets.
To make searches of this sort successful, it will be necessary to
(1) improve the duty cycle and coverage of the observations over that
possible from a single observing site,
(2) attain the best possible precision in the photometric time series, and
(3) plan to observe more stars, by factors of several, than has been
undertaken to date.

Spaceborne missions searching for transits by Earth-sized planets will
need to take care in the selection of their target lists, so that
resources are not wasted on unsuitable, large-radius stars.
The problem of false alarms will be less serious for these missions,
except for diluted transits arising from faint background eclipsing
binary stars.
For both ground- and space-based observations, it would be useful to
extend the results of the present modeling effort to regions off the
galactic plane.

\acknowledgments
I am grateful to David Latham and Didier Queloz 
for helpful conversations, and to Robert Noyes and the anonymous referee
for careful readings of the paper's initial versions.
The National Center for Atmospheric Research is sponsored
by the National Science Foundation.
This publication makes use of data products from the Two Micron All Sky Survey,
which is a joint project of the University of Massachusetts and the Infrared
Processing and Analysis Center/California Institute of Technology, 
funded by the National Aeronautics and Space Administration 
and the National Science Foundation.

\clearpage

\figcaption{
Cumulative distribution of the periods of known extrasolar planets
(solid line) from the tabulation in \citet{sch03}.
Also shown is a piecewise linear fit (heavy dashed line).
The derivative of the fitted curve gives the probability density
$P(\log \Pi_p)$ quoted in the text.
}

\figcaption{
Histograms of star density vs. J-K color, in units of stars per square
degree per .05-magnitude bin, in a sample with R$\leq$12.
Thick solid curve: observations from the 2MASS catalog, 
for the STARE Cygnus field.
Thin solid curve: model, for main-sequence stars only.
Dashed curve: model, for giants only.
Hatched area: stars with radii less than 1.3 solar radii.
}

\figcaption{
Marginal probabilities for bins in $\log \delta$, for the cases
MSU (solid), MPU (dashed), MSDF (dot-dashed), and MSDT (triple dot-dashed).
The hatched area shows those MPU systems with depths exceeding 1\%.
All are estimated assuming a survey that is magnitude-limited at R=12,
with no deleterious effects from the observing window function.
}

\clearpage

\begin{deluxetable}{lcccc}
\tablewidth{0pt}
\tablecaption{Detection Rates per $10^4$ Stars}
\tablehead{
\colhead{Category} & {$n \geq 0$} & {$n \geq 2$} & {$n \geq 3$} &\qquad Kepler}
\startdata
MPU &1.43 &0.74 &0.39 &\qquad 0.004 \\
MSU &4.56 &2.92 &2.27 &\qquad 0.01 \\
MSDF &1.90 &1.52 &1.26 &\qquad 28.60 \\
MSDT &1.64 &1.20 &0.98 &\qquad 0.23 \\
GSU &0.00 &0.00 &0.00 & \qquad 0.81 \\
\enddata
\end{deluxetable}
\clearpage

\begin{figure}
\plotone{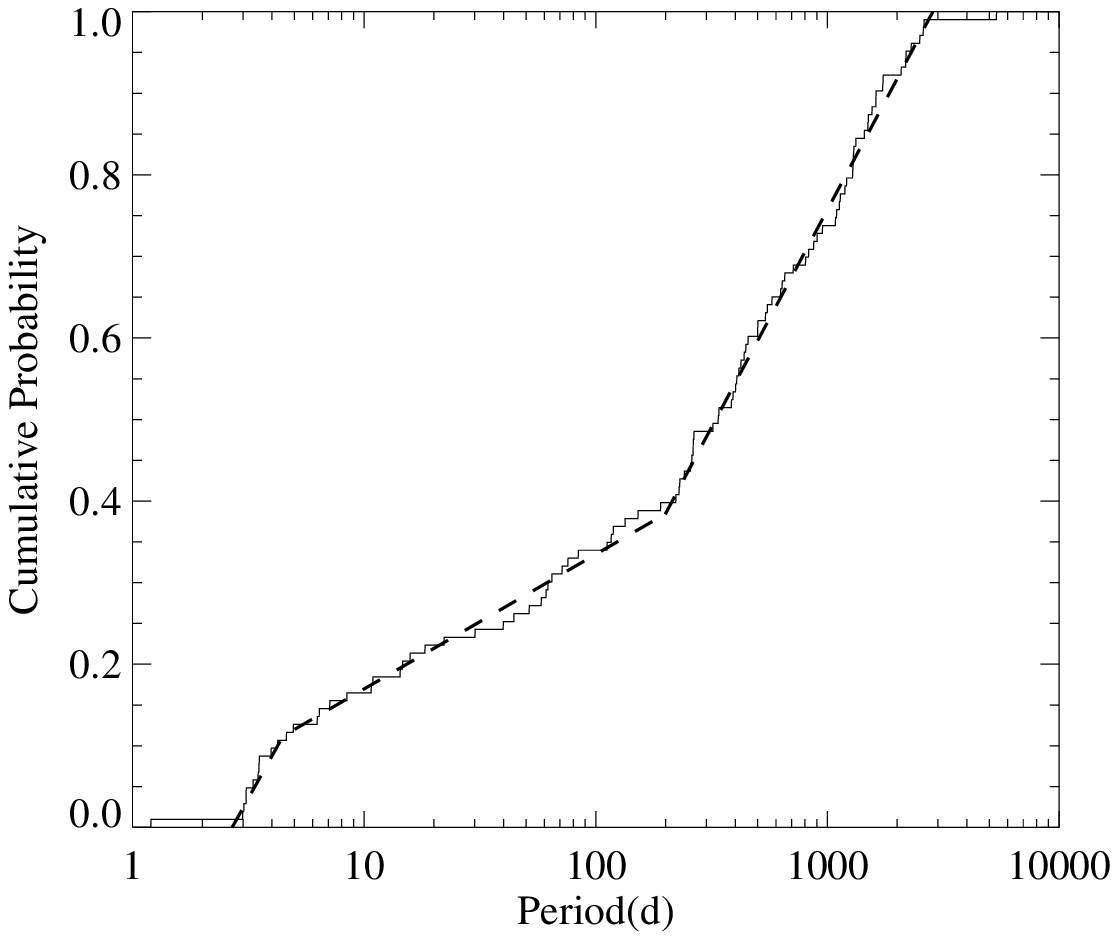}
\end{figure}

\begin{figure}
\plotone{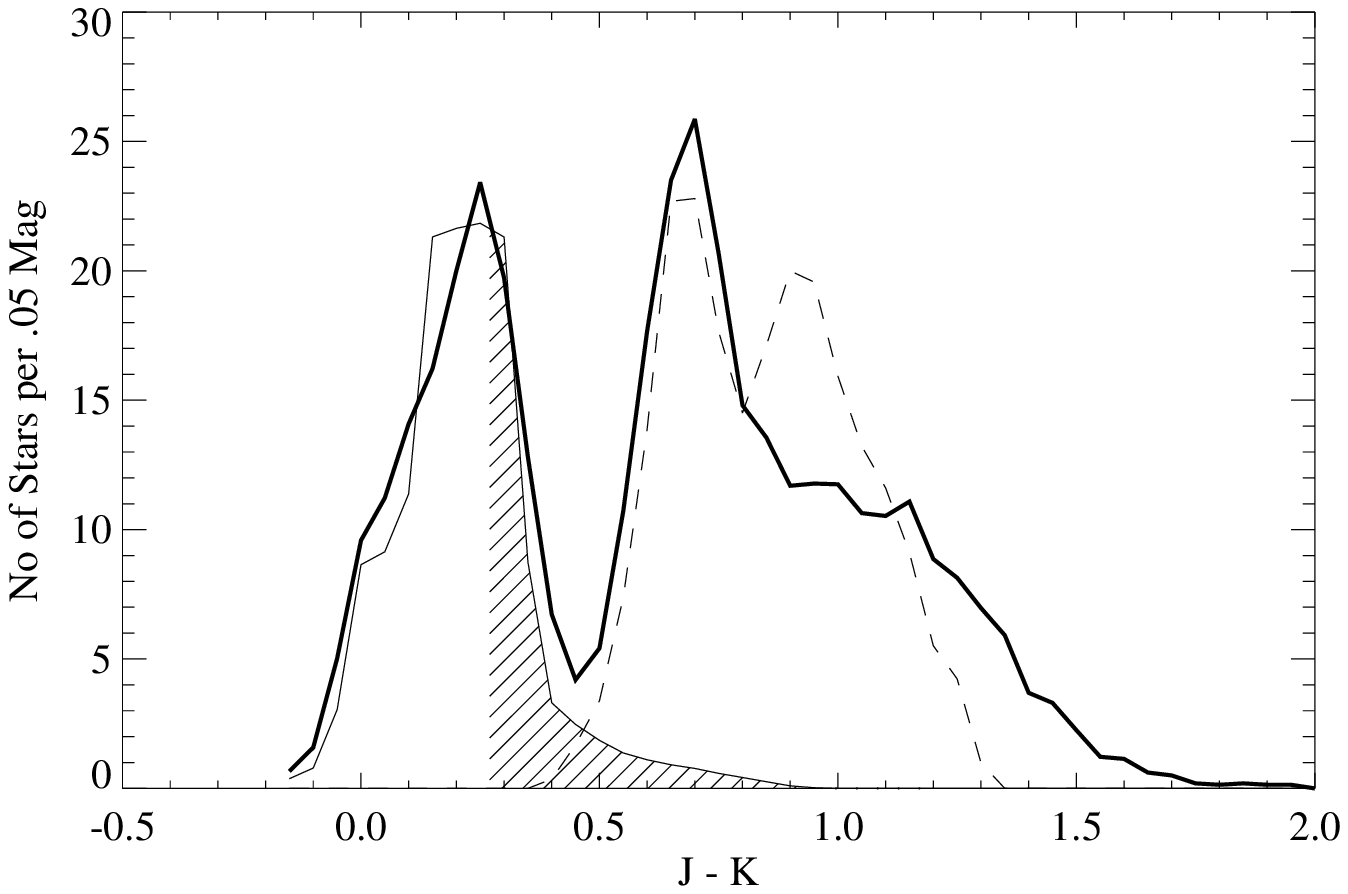}
\end{figure}

\begin{figure}
\plotone{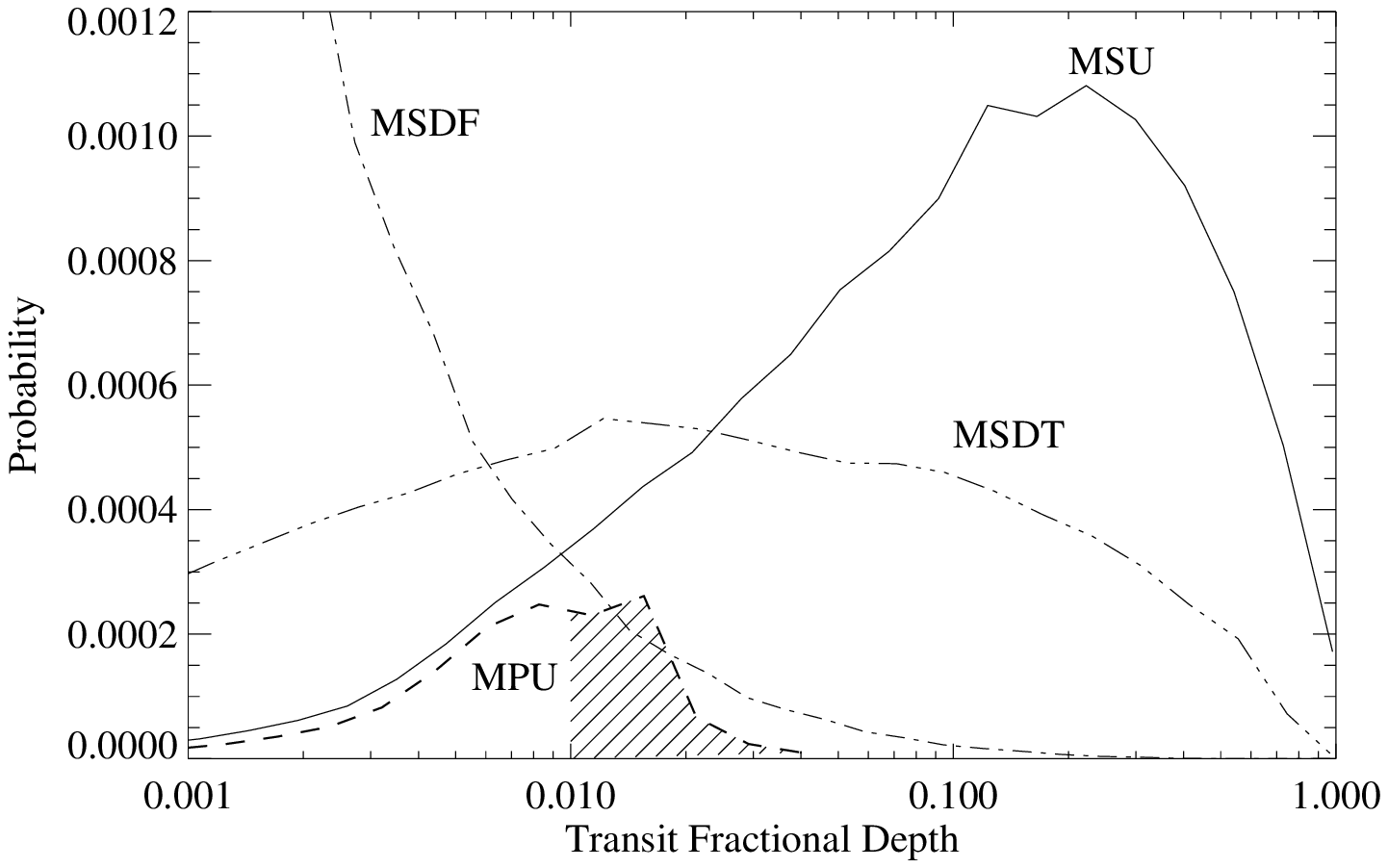}
\end{figure}

\end{document}